\documentclass[aps,prl,twocolumn,superscriptaddress,longbibliography]{revtex4-2}
\usepackage{graphicx}
\usepackage{amssymb,amsmath}
\usepackage{bm}
\usepackage{dcolumn}
\usepackage{float}
\usepackage[OT1]{fontenc} 
\usepackage{url}
\usepackage{mathrsfs}
\usepackage{slashed,comment}
\usepackage{color}
\usepackage{verbatim}
\usepackage{enumitem}
\usepackage{soul,physics}
\usepackage[driverfallback=dvipdfm]{hyperref}
\hypersetup{pdfpagemode=FullScreen,colorlinks=true,breaklinks,urlcolor=blue,linkcolor=blue,citecolor=blue}

\usepackage{subfigure}
\usepackage{amssymb}
\usepackage{bm}
\usepackage{graphicx}
\usepackage{amsmath,amssymb}
\usepackage{tikz,fp}
\usepackage{tikz-cd}
\usetikzlibrary{arrows}
\usetikzlibrary{intersections}
\usetikzlibrary{shapes.geometric}
\usetikzlibrary{decorations.pathmorphing, patterns,shapes,fixedpointarithmetic}
\usetikzlibrary{decorations.markings}

\usepackage{pdfpages} 
\usepackage{pgffor} 

\makeatletter
\AtBeginDocument{\let\LS@rot\@undefined}
\makeatother

\def\supplementfilename{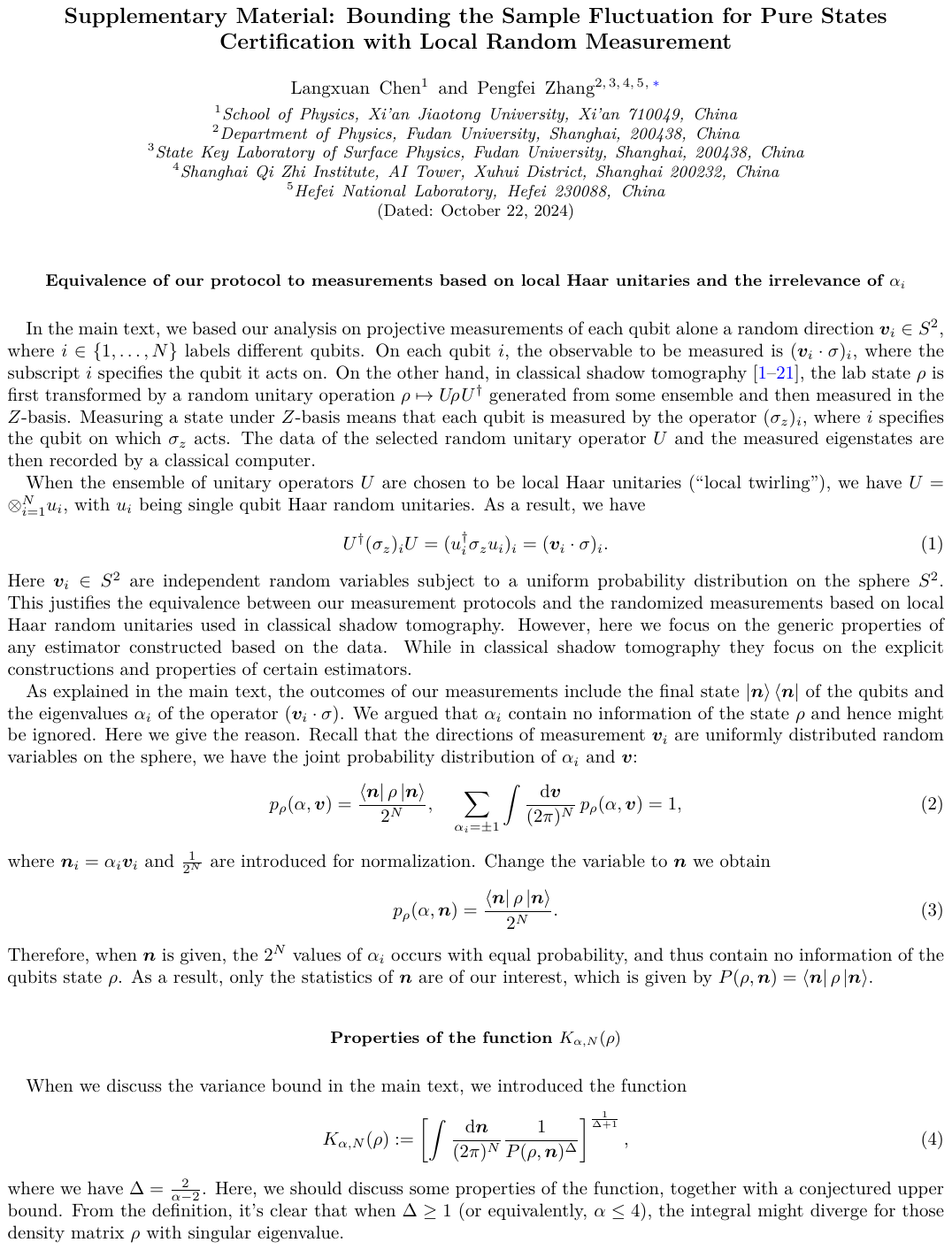}

\pdfximage{\supplementfilename}
\def\numbersupplementpages{\the\pdflastximagepages}

\newif\ifarXiv
\arXivtrue 

\begin{document}
  
  \title{Bounding the Sample Fluctuation for Pure States Certification \\with Local Random Measurement}
  
  \author{Langxuan Chen}
  \affiliation{School of Physics, Xi'an Jiaotong University, Xi'an 710049, China}

  \author{Pengfei Zhang}
  \thanks{PengfeiZhang.physics@gmail.com}
  \affiliation{Department of Physics, Fudan University, Shanghai, 200438, China}
  \affiliation{State Key Laboratory of Surface Physics, Fudan University, Shanghai, 200438, China}
  \affiliation{Shanghai Qi Zhi Institute, AI Tower, Xuhui District, Shanghai 200232, China}
  \affiliation{Hefei National Laboratory, Hefei 230088, China}
  \date{\today}

  \begin{abstract}
 Remarkable breakthroughs in quantum science and technology are demanding for more efficient methods in analyzing quantum many-body states. A significant challenge in this field is to verify whether a quantum state prepared by quantum devices in the lab accurately matches the desired target pure state. Recent advancements in randomized measurement techniques have provided fresh insights in this area. Specifically, protocols such as classical shadow tomography and shadow overlap have been proposed. Building on these developments, we investigate the fundamental properties of schemes that certify pure quantum states through random local Haar measurements. We derive bounds for sample fluctuations that are applicable regardless of the specific estimator construction. These bounds depend on the operator size distribution of either the observable used to estimate fidelity or the valid variation of the reduced density matrix for arbitrary observables. Our results unveil the intrinsic interplay between operator complexity and the efficiency of quantum algorithms, serving as an obstacle to local certification of pure states with long-range entanglement.
  \end{abstract}
  
  \maketitle

      \begin{figure}[t]
    \centering
    \includegraphics[width=0.95\linewidth]{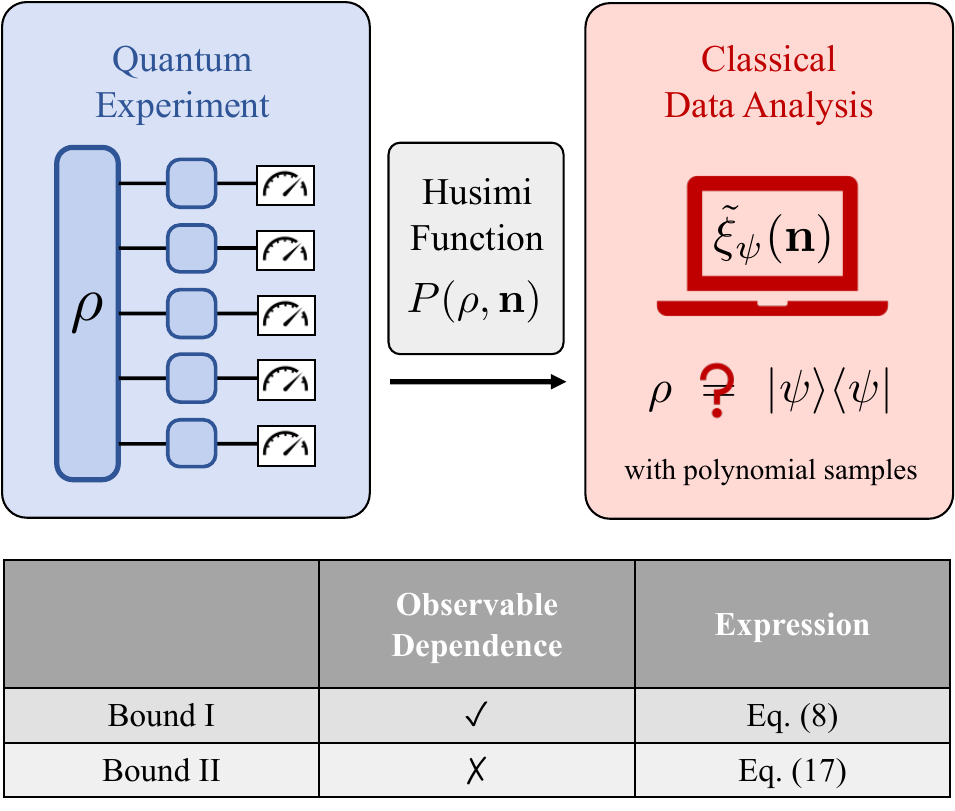}
    \caption{Schematic of our setup for the certification of pure states using random local Haar measurements. In this setup, the lab state is measured in the computational basis following random single-qubit rotations, which adhere to the probability distribution described by the Husimi function. We investigate whether the data can be used to efficiently certify pure states by deriving bounds on the sample fluctuation. The main results are given by Eq.~\eqref{eqn:res_gen} or Eq.~\eqref{eqn:main2}.}
    \label{fig:schemticas}
  \end{figure}

  \emph{ \color{blue}Introduction.--} Recent years have witnessed rapid advancements in quantum science and technology~\cite{Preskill:2018jim}. State-of-the-art experimental platforms, such as Rydberg atom arrays~\cite{Evered:2023wbk,Ma:2023ltx,Bluvstein:2023zmt,2019arXiv190407369B,doi:10.1126/science.abg2530,Ebadi:2022oxd,Bluvstein:2021jsq,Lis:2023gaz,Manetsch:2024lwl,Tao:2023uxy,Cao:2024dhc}, have enabled the development of quantum processors with hundreds of qubits. This marks a new era in quantum computation, allowing for efficient simulation and optimization of complex systems, including those that arise in molecular dynamics, materials science, and cryptography that are well beyond the capabilities of classical methods. However, a significant challenge remains: the efficient characterization of quantum many-body states, which often demands computational resources that scale exponentially in system size, making the process difficult to control and optimize. One concrete example of such challenge is the certification of quantum states, which requires determining whether a laboratory-prepared state matches the target state within certain accuracy~\cite{Hayashi_2006, Cramer2010, Aolita2015, Lanyon2017, PRXQuantum.2.010201, PhysRevA.99.052346, PhysRevA.100.032316, PhysRevLett.123.260504, Eisert2020, PRXQuantum.2.010102, Yu2021StatisticalMF, PRXQuantum.3.010317, RevModPhys.95.035001, https://doi.org/10.1002/qute.202100118, Zhu_2024, Huang:2024qmc}.

  The randomized measurement toolbox has emerged as a promising strategy for characterizing quantum states~\cite{PhysRevA.99.052323, doi:10.1126/science.aau4963, Elben:2022jvo}. Compared to traditional algorithms, it involves applying random unitary transformations drawn from a pre-chosen ensemble before performing measurements in the computational basis. A well-established example of this approach is classical shadow tomography~\cite{Aaronson2017ShadowTO, Huang:2020tih, paini2019approximatedescriptionquantumstates, Zhou2024efficientclassical, PhysRevLett.127.110504, Ippoliti:2022vfn, PRXQuantum.2.010307, PRXQuantum.2.030348, doi:10.1126/science.abk3333, PhysRevResearch.4.013054, PhysRevResearch.6.013029, PhysRevB.109.094209, PRXQuantum.5.010350, PRXQuantum.5.020304, PhysRevResearch.5.023027, Akhtar2023scalableflexible, akhtar2024, 2024arXiv240217911H, 2024arXiv240611788Z, PhysRevLett.133.020602, Bu:2024aa}, which enables the prediction of many properties with very few measurements. However, predicting fidelity with low sample complexity requires applying Global Haar random unitaries to the entire system, leading to high circuit complexity. On the other hand, the sample complexity of protocols employing single-qubit Haar random unitaries grow exponentially for generic quantum states, making them impractical for realistic experiments. Later, the shadow overlap was proposed as an alternative approach to overcome this difficulty~\cite{Huang:2024qmc}, which only relies on single-qubit measurement and can certify almost all quantum states within polynomial time. 

 In this work, instead of focusing on specific protocols, we investigate the intrinsic bounds on sample complexity when only local single-qubit Haar random measurements are accessible. We find that for a fixed observable used for certification, the sample fluctuations are bounded by the generating function of the observable's size distribution. This provides insight into how certain protocols can outperform others, as exemplified by direct fidelity estimation and shadow overlap. We further derive a bound that is independent of any specific observable, which presents an obstacle for arbitrary protocols in quantum state certification based on local Haar measurements. For example, we demonstrate that GHZ-like states cannot be efficiently certified through local Haar measurements, thus ruling out the possibility of constructing a generic (local) protocol that can certify all quantum states. Our results provide a direct correspondence between the operator complexity and the efficiency of quantum algorithms.

  \emph{ \color{blue}Setup.--} We consider the preparation of a target pure state $\ket{\psi}$ on quantum devices. Due to the potential presence of noise, we need to know whether the laboratory density matrix $\rho$ matches the target pure state $\ket{\psi}$. A standard measure for this is the fidelity $F_\psi(\rho)=\bra{\psi}\rho\ket{\psi}$. More generally, we can construct certain observable that provides information of the fidelity, such as the shadow overlap~\cite{Huang:2024qmc}. The criteria for selecting a particular operator that can be predicted efficiently are therefore of special interest.

  In this work, we focus on a class of protocols utilizing the random measurement toolbox~\cite{Elben:2022jvo}, which eliminate local basis dependence.  In our measurement procedure, each qubit (labeled by $i\in\{1,\ldots,N\}$) is measured in the eigenbasis of the observable $(\bm{v}_{i}\cdot \sigma)$, where $\bm{v}_i\in S^2$ is a randomly selected direction of measurement for the $i$-th qubit. The outcomes of measurements are described by the measured eigenvalues $\alpha_{i}=\pm 1$ (i.e. ``up'' or ``down'') of the operators $(\bm{v}_{i}\cdot \sigma)$, and the final state $\ket{\bm{n}}\bra{\bm{n}}$ of the qubits. Here $\ket{\bm{n}}=\ket{\bm{n}_1}\otimes \ket{\bm{n}_2}...\ket{\bm{n}_N}$ is a product of spin coherent states with $\bm{n}_{i}=\alpha_{i}\bm{v}_i $. As elaborated in~\cite{Xu:2023chv}, the statistics of the final vector $\bm{n}$ are described by the Husimi function $P(\rho,\bm{n})\equiv  \bra{\bm{n}}\rho\ket{\bm{n}}$~\cite{1940264},  which is normalized as $\int \frac{d\bm{n}}{(2\pi)^{N}}~P(\rho,\bm{n})=1$.  On the other hand, the eigenvalues $\alpha_i$ carries no information and can thus be ignored. We should note that the measurement described here is equivalent to the measurements performed in classical shadow tomography employing single qubit Haar random unitaries. Details about the irrelevance of $\alpha_i$ and the connection with measurements based on local Haar unitaries are discussed in the supplementary material~\cite{SM}. 
Note that the measurement protocol here is independent of the state $\ket{\psi}$ to be certified (or more general properties to be investigated), that is, we follow the philosophy of ``\textit{measure first, ask question later}''~\cite{Elben:2022jvo}.

  An unbiased measurement of an observable $\xi(\rho)=\text{tr}[\hat{\xi}\rho]$ based on the measured values of $\bm n$ might be obtained by taking the sample average of some estimator $\tilde{\xi}(\bm{n})$, which is a classical function of the measurement outcome $\bm{n}$:
  \begin{equation}\label{eqn:estimator}
   \mathbb{E}[\tilde{\xi}]=  \int \frac{\mathrm{d}\bm{n}}{(2\pi)^{N}}~P(\rho,\bm{n})~\tilde{\xi}(\bm{n})=\text{tr}[\hat{\xi}\rho].
  \end{equation}
  As an example, in the standard classical shadow tomography, the estimator is constructed by noticing that the expectation of the classical snap shot $\sigma_{\bm{n}}=\ket{\bm{n}}\bra{\bm{n}}$ defines a measurement channel $\mathcal{M}$ on the density matrix, that is, $ \mathbb{E}[\sigma_{\bm{n}}]=\mathcal{M}\!\left(\rho\right)$. The classical estimator is then constructed as $\tilde{\xi}(\bm{n})=\text{tr}\left[\mathcal{M}^{-1}\!\left(\ket{\bm{n}}\!\bra{\bm{n}}\right)\hat{\xi}\,\right]$, which satisfies $\mathbb{E}[\tilde{\xi}(\bm n)]=\text{tr}[\hat{\xi}\rho]$. Note that while estimating any observable $\hat{\xi}$ may be of interest, we will focus on those that help certify certain quantum states $\psi$, denoted by $\hat{\xi}_{\psi}$. Nevertheless, the formalism developed here are also apply for the estimation of generic observable $\hat{\xi}$ without much change. 

  \emph{ \color{blue}Sample Fluctuation and Size.--} A generic certification scheme depends on both $\hat{\xi}_\psi$ and $\tilde{\xi}_{\psi}(\bm{n})$: while different choices of estimators may yield the same $\hat{\xi}_\psi$, their variances may differ significantly, leading to variations in sample complexity. However, we demonstrate that there is an obstacle to constructing efficient estimators following a few steps: 

  \textbf{Step 1}. We consider a variation of Eq.~\eqref{eqn:estimator} for a small change $\delta \rho$ of the density matrix $\rho$:
  \begin{equation}\label{eqn:variation}
  \mathbb{E}[(\delta P/P)\times (\tilde{\xi}_\psi-\xi_\psi)]=\text{tr}[\hat{\xi}_\psi\delta \rho].
  \end{equation}
  Here, we introduce $\delta P(\bm{n}):=P(\delta\rho,\bm{n})$ and use the relation $\mathbb{E}[\delta P/P]=0$ due to the traceless condition $\text{tr}[\delta \rho]=0$. Although the original derivation requires $\rho+\delta \rho$ to be positive semidefinite, Eq.~\eqref{eqn:variation} is linear in $\delta \rho$, making it valid for any traceless operator $\delta \rho$ by linear superposition. For clarity, we now replace $\delta \rho$ with $O$ to avoid potential confusion.

  \textbf{Step 2}. We now apply generalized H{\"o}lder's inequality $\rVert fgh\rVert_1\leq\rVert f\rVert_\alpha \rVert g\rVert_\beta \rVert h\rVert_\gamma$ when $\alpha,\beta,\gamma\geq 1$ and $\frac{1}{\alpha}+\frac{1}{\beta}+\frac{1}{\gamma}=1$. Here, we introduce the $\alpha$-norm ($\alpha\geq1$) with respect to the Husimi function, defined as $\rVert f\rVert_{\alpha}:=\mathbb{E}\left[|f|^{\alpha}\right]^{1/\alpha}= \left[\int \frac{d\bm{n}}{(2\pi)^{N}} P(\rho,\bm{n}) |f(\bm{n})|^\alpha\right]^{1/\alpha}.$ By setting $f=\tilde{\xi}_\psi-\xi_\psi$, $g=\delta P/\sqrt{P}$, $h=1/\sqrt{P}$ and $\beta=2$ we obtain 
   \begin{equation}\label{eqn:main1}
  \left\rVert \tilde{\xi}_\psi-\xi_\psi\right\rVert_{\alpha}\geq \max_{\text{tr}[O]=0}\frac{\left|\text{tr}[\hat{\xi}_\psi O]\right|}{\sqrt{K_{\alpha,N}(\rho)\times\left[\int \frac{\mathrm{d}\bm{n}}{(2\pi)^{N}}~P(O,\bm{n})^{2}\right]}}.
  \end{equation}
Here, the L.H.S. measures the sample fluctuations of $\tilde{\xi}_\psi(\bm{n})$. Note that we must have $\alpha\geq 2$ in the inequality. On the R.H.S., we have introduced a maximization over arbitrary traceless Hermitian operator $O$, and for conciseness, we have defined
  \begin{equation}
  K_{\alpha,N}(\rho):=\left[\int \frac{\mathrm{d}\bm{n}}{(2\pi)^{N}}\frac{1}{P(\rho,\bm{n})^\Delta}\right]^{\frac{\alpha-2}{\alpha}},
  \end{equation}
  with $\Delta=\frac{\gamma}{2}-1=\frac{2}{\alpha-2}$. The properties of $K_{\alpha,N}(\rho)$ are discussed in detail in the supplementary material~\cite{SM}, together with a conjectured bound. There are two regimes when $K_{\alpha,N}(\rho)$ is easy to calculate. First, if we maximize the R.H.S. of Eq.~\eqref{eqn:main1} over all possible lab state $\rho$, then $K_{\alpha,N}(\rho)$ should be minimized. It was shown in the supplementary material~\cite{SM} that $K_{\alpha,N}(\rho)$ is always minimized by $\rho_0=I/2^{N}$, for which $K_{\alpha,N}(\rho_0)=2^{N}$. On the other hand, if we are only interested in the case when $\alpha$ is large (but finite) then $\Delta \approx 0$, leading to $K_{\alpha,N}(\rho)\approx 2^{N}$ for any choice of $\rho$. In either regime we have the following estimate:
  \begin{equation}
  \label{eqn:main1:1}
  \left\rVert \tilde{\xi}_\psi-\xi_\psi\right\rVert_{\alpha}\geq \max_{\text{tr}[O]=0}\frac{\left|\text{tr}[\hat{\xi}_\psi O]\right|}{\sqrt{2^{N}\times\left[\int \frac{\mathrm{d}\bm{n}}{(2\pi)^{N}}~P(O,\bm{n})^{2}\right]}}.
  \end{equation}

  \textbf{Step 3}. Finally, the maximization over $O$ can be worked out explicitly. We first perform the integral over $\bm{n}$ in the denominator. Similar integrals appear when computing the shadow norm of classical shadow tomography~\cite{Huang:2020tih} or second R\'enyi Wehrl entropy~\cite{newpaper}. Expanding both the measurement operator $\hat{\xi}_\psi$ and $O$ in the basis of Pauli strings $\{P\}$ as
  \begin{equation}
  \hat{\xi}_\psi=\sqrt{\langle\hat{\xi}_\psi^2\rangle}\sum_Pc_\xi(P)P,\ \ \ \ \ \ O=\sum_Pc_O(P)P,
  \end{equation} 
  where we introduced $\langle\hat{\xi}_\psi^2\rangle\equiv\text{tr}[\hat{\xi}_\psi^2]/2^N$ and $c_\xi(P)$ satisfies the normalization condition $\sum_Pc_\xi(P)^2=1$. Eq.~\eqref{eqn:main1:1} becomes
  \begin{equation}
  \left\rVert \tilde{\xi}_\psi-\xi_\psi\right\rVert_{\alpha}\geq \sqrt{\langle\hat{\xi}_\psi^2\rangle}\max_{O}\frac{\left|\sum_{P}c_\xi(P)c_O(P)\right|}{\sqrt{\sum_P 3^{-s(P)} c_O(P)^2}
  }.
  \end{equation}
  Here, the operator size $s(P)$ counts the number of non-trivial Pauli operators in the Pauli string operator $P$. An direct maximization leads to $c_O(P)=3^{s(P)}\times c_\xi(P)$ for $P\neq I$, while $c_O(I)=0$ due to the traceless condition. As a result, we find that $\left\rVert \tilde{\xi}_\psi-\xi_\psi\right\rVert_{\alpha}^2\geq \langle\hat{\xi}_\psi^2\rangle\sum_{s>0} 3^s P_{\xi}(s)$, where $\mathcal{P}_\xi(s):=\sum_{P|s(P)=s}c_\xi(P)^2$ is known as the operator size distribution of $\hat{\xi}$, and is currently an active subject of research~\cite{Nahum:2017yvy, Qi:2018bje, Hunter-Jones:2018otn, vonKeyserlingk:2017dyr, Khemani:2017nda, Dias:2021ncd, PhysRevResearch.3.L032057, Roberts:2015aa, Roberts:2018aa, qi2019, Lucas:2020pgj, Lensky:2020ubw, PhysRevLett.122.216601, Chen:2019klo, Chen:2020bmq, Yin:2020oze, Zhou:2021syv, Omanakuttan:2022ikz, Ippoliti:2022vfn, Bu:2024aa, Zhang2023, liu2023signaturescrambloneffectivefield, PhysRevLett.130.250401, Zhang2024,PhysRevB.110.035137}. 
  Introducing the notation 
  \begin{equation}
  \overline{3^{s}}:= \sum_{P} 3^{s(P)}c_\xi(P)^2= \sum_{s}3^s\mathcal{P}_\xi(s),
  \end{equation}
  we have 
 \begin{equation}\label{eqn:res_gen}
  \left\rVert \tilde{\xi}_\psi-\xi_\psi\right\rVert_{\alpha}\geq \sqrt{\langle\hat{\xi}_\psi^2\rangle(\overline{3^s}-1)}.
  \end{equation}

  Eq.~\eqref{eqn:res_gen} is one of our main results. This inequality only relies on the observable $\hat{\xi}_\psi$, and is valid for arbitrary estimator $\tilde{\xi}_\psi(\bm{n})$. It demonstrates that the sample fluctuation is bounded by a generating function of the operator size distribution, which serves as a quantitative criterion for comparing different protocols: If we consider a number of samples $\{\bm{n}^{(1)},\bm{n}^{(2)},...,\bm{n}^{(M)}\}$ independently drawn from the distribution $P(\rho,\bm{n})$, the fluctuation of the sample average satisfies $|\sum_i\tilde{\xi}_\psi(\bm{n}^{(i)})/M-\xi_\psi|^\alpha\approx \Big(\left\rVert \tilde{\xi}_\psi-\xi_\psi\right\rVert_{\alpha}\Big)^\alpha/M^{\alpha-1}$ when $\alpha$ is an even integer. Here, we have neglected all odd moments. Therefore, when the R.H.S. of Eq.~\eqref{eqn:main1} becomes exponential in $N$, the prediction of $\xi_\psi$ requires exponentially large sample size $M$. Details about the relations between the bound and the sample size are discussed in the supplementary material~\cite{SM}. 
  Below, we present a few examples using concrete protocols and experimental relevant states. A related inequality, independent of $\hat{\xi}_\psi$, will be discussed subsequently. Note that (as we have mentioned in the set up), although our focus is on the certification of pure states $\psi$, the inequality above also provides a bound for the measurement of any observable $\hat{\xi}$, as explained in the supplementary material~\cite{SM}.
  
  \emph{ \color{blue}Example 1: Direct Fidelity Estimation.--} Our first example concerns the direct measurement of the fidelity (see e.g.~\cite{PhysRevLett.106.230501, PhysRevLett.107.210404}), which corresponds to choosing the observable $\hat{\xi}_{\psi}$ to be $\hat{F}_\psi=P_\psi=\ket{\psi}\bra{\psi}$ and therefore $\langle (\hat{F}_\psi)^2\rangle =1/2^N$. We examine the bound~\eqref{eqn:res_gen} for Haar random states, which are of particular interests. For a typical Haar random state, it is reasonable to approximate the size distribution by its expectation value over the Haar ensemble. For $P\neq I$, this gives $c_F (P)^2=\frac{1}{2^N(2^N+1)}$ and the size distribution is a binomial distribution
  \begin{equation}
  \mathcal{P}_{\text{Haar}}^{F}(s)={N\choose s}~\frac{3^s}{2^N(2^N+1)},\ \ \ \ \ \ \text{for }s>0. 
  \end{equation} 
 For large $\alpha$, It leads to the bound $\left\rVert \tilde{F}_\psi-F_\psi\right\rVert_{\alpha}\geq(\sqrt{5/4})^N$, which grows exponentially in the limit $N\rightarrow \infty$. Therefore, a direct measurement of the fidelity based on random local Haar measurement always requires exponential many samples for Haar random states. The results are intuitively consistent with our expectation that local measurements, which are inherently inefficient at characterizing complex quantum entanglement, will fail to accurately predict the fidelity of generic quantum states.

  \emph{ \color{blue}Example 2: Randomized Shadow Overlap.--} Now, we turn our attention to the shadow overlap, a protocol proposed to certify nearly all quantum states using only a few single-qubit measurements. The original scheme requires measuring all but one qubit in the computational basis and performing classical shadow tomography on the remaining qubit. A brief review of this approach is provided in the supplementary material~\cite{SM}. The protocol corresponds to measuring the observable
  \begin{equation}\label{eqn:L}
  \hat{L}_\psi=\frac{1}{N}\sum_{k=1}^N\sum_{\{z_i\}\setminus z_k}\ket{\{z_i\}}\bra{\{z_i\}}\otimes \frac{\langle \{z_i\}|\psi\rangle \langle \psi|\{z_i\}\rangle}{\langle \psi|\{z_i\}\rangle \langle \{z_i\}|\psi\rangle }.
  \end{equation} 
  Here, $\{z_i\}\!\setminus \!z_k$ denotes the summation over all $z_i\in\{0,1\}$ with $i\neq k$. Consequently, $|\{z_i\}\rangle$ represents a state defined on the Hilbert space of $N-1$ qubits and $\langle \{z_i\}|\psi\rangle \langle \psi|\{z_i\}\rangle$ is an operator acting solely on site $k$. It is straightforward to verify that $\bra{\psi}\!\hat{L}_\psi\!\ket{\psi}=1$. Furthermore, it was shown that $\hat{L}_\psi$ can be used for the efficient certification of almost all quantum states~\cite{Huang:2024qmc} provided that its spectrum exhibits a gap that is not exponentially small in $N$. We offer an explanation for this efficiency by analyzing the operator size of the measurement operator in a generalized scheme with random local Haar measurements on every qubit. This leads to a random-averaged observable $\hat{\xi}_\psi=\hat{\Omega}_\psi$ with
  \begin{equation}
  \hat{\Omega}_\psi=\overline{U^\dagger\hat{L}_{U\hspace{-0.1em}\psi}\,U}=\int_{\text{Haar}} \prod_i \mathrm{d}u_i\, U^\dagger\hat{L}_{U\hspace{-0.1em}\psi}\,U.
  \end{equation}
  Here, $U=\otimes_{i=1}^N u_i$ and $u_i$ is a single-qubit random Haar unitary for site $i$. More details of $\hat{\Omega}_\psi$ are provided in the supplementary material~\cite{SM}. 

  Now, we examine our bound Eq.~\eqref{eqn:main1} for the randomized shadow overlap. The R.H.S. of the bound is determined by the function 
  \begin{equation}
  t_{\psi}=\langle\hat{\Omega}^2_{\psi}\rangle\sum_{s>0} 3^s P_{\Omega}(s).
  \end{equation} The numerical results for $t_{\psi}$ are shown in table~\ref{tab:table1} below, where the sample average $\bar{t}$ and the sample standard deviation $S_t$ are evaluated for randomly generated Haar states $\{\psi_{i}\}_{i=1}^M$ with sample size $M$. It's clear that $\bar{t}$ are bounded for Haar random states regardless of the system size $N$.
  \begin{table}[b]
  \caption{\label{tab:table1}
  Statistics of the function $t_{\psi}$ for randomly generated Haar states $\psi$ with system size $N$ and sample size $M$
}
\begin{ruledtabular}
\begin{tabular}{lccr}
$N$&
$\bar{t}$&
$S_t$&
$M$\\
\colrule
2 & 0.5183 & 0.1105 & 1000\\ 
3 & 0.4271 & 0.0514 & 500\\
4 & 0.3820 & 0.0258 & 200 \\
5 & 0.3528 & 0.0134 & 100 \\
6 & 0.3331 & 0.0072 & 100\\
7 & 0.3186 & 0.0047 & 100\\
\end{tabular}
\end{ruledtabular}
\end{table}

Although we are not able to obtain analytical results, we provide a qualitative understanding of how the operator size of $\hat{\Omega}_\psi$ becomes significantly smaller than that of $\hat{F}_\psi$. For typical Haar random states, we approximate $\hat{L}_{\psi}$ by taking the ensemble average over $|\psi\rangle$ independently for both $\langle \{z_i\}|\psi\rangle \langle \psi|\{z_i\}\rangle$ and $\langle \psi|\{z_i\}\rangle \langle \{z_i\}|\psi\rangle$. The result reads
  \begin{equation}
  \hat{\Omega}_\psi=\overline{\hat{L}}\approx \frac{1}{2N}\sum_{k=1}^N\sum_{\{z_i\}/z_k}\ket{\{z_i\}}\bra{\{z_i\}}\otimes \hat{I}_k=\frac{\hat{I}}{2}.
  \end{equation}
  Here, we have used the fact that the single-qubit Haar random unitary can be neglected for Haar random states. As a result, the operator size is exactly zero and does not impose any meaningful bound on $\left\rVert \tilde{\Omega}_\psi-\Omega_{\psi}\right\rVert_{\alpha}$. When the fluctuations of $\langle \{z_i\}|\psi\rangle \langle \psi|\{z_i\}\rangle$ or $\langle \psi|\{z_i\}\rangle \langle \{z_i\}|\psi\rangle$ are taken into account, a small but finite size for $\hat{\Omega}_\psi$ emerges.

  \emph{ \color{blue}Observable Independent Bound.--} The above results show that using different observables (and estimators) may lead to significant improvements in fidelity estimation. It is then natural to ask whether there exists protocols based on local Haar measurements that effectively certifies \textit{all} quantum states by a clever choice of the observable. Here, we prove another bound that serves as an obstacle to the local efficient certification procedure, regardless of the choice of observable $\hat{\xi}_\psi$. To achieve observable independence, note that although we may use any $\hat{\xi}_{\psi}$, the expectation value $\xi_{\psi}(\rho)=\text{tr}[\hat{\xi}_\psi\rho]$ must tell us whether the fidelity is close to one or not. Without loss of generality, we may assume that $\xi_{\psi}(\rho)\leq 1$, $\bra{\psi} \hat{\xi}_{\psi}\ket{\psi}=1$, and most importantly, we shall require
  \begin{equation}
  \label{eqn:wittness1}
      1-\xi_{\psi}(\rho)\geq \frac{1}{\tau}(1-F_{\psi}(\rho)),
  \end{equation}
  where $\tau$ is a proportionality constant, and might be set to 1 with a redefinition of $\xi$. Note that when Eq.~\eqref{eqn:wittness1} fails, there are lab states $\rho$ with $\xi_{\psi}(\rho)=1$ but $F_{\psi}(\rho)<1$, results in failure of the certification. Estimators that do not satisfy Eq.~\eqref{eqn:wittness1} are therefore unable of  certifying the state $\ket{\psi}$. 
 
 Now, let us consider lab states $\rho$ that are close to the target state $P_{\psi}=\ket{\psi}\!\bra{\psi}$, that is, $\rho=\ket{\psi}\!\bra{\psi}+\delta\rho$. In this case, we have 
  \begin{equation}\label{eqn:wittness}
  \left|\xi_{\psi}(\delta\rho)\right| \geq \frac{1}{\tau}\left|\text{tr}[P_\psi\delta \rho]\right|.
  \end{equation} 
  Now, we attempt to follow the steps from the previous section to derive a bound based on Eq.~\eqref{eqn:wittness}. The key difference now is that the positivity requirement $P_\psi+\delta \rho\geq 0$ can no longer be lifted since Eq.~\eqref{eqn:wittness} is not a linear equality. Consequently, we find that
  \begin{equation}
  \label{eqn:main2}
  \tau\left\rVert \tilde{\xi}_\psi-\xi_\psi\right\rVert_{\alpha}\geq\max_{P_\psi+\delta\rho\geq 0}\frac{\left|\text{tr}[P_{\psi}\delta\rho]\right|}{\sqrt{K_{\alpha,N}(P_{\psi})\left[\int \frac{\mathrm{d}\bm{n}}{(2\pi)^{N}}~\bra{\bm n}\delta\rho\ket{\bm n}^2\right]}},
  \end{equation}
The combination $\tau\left\rVert \tilde{\xi}_\psi-\xi_\psi\right\rVert_{\alpha}$ on the L.H.S. appears since to distinguish high-fidelity and low-fidelity lab states $\rho$ one have to control the uncertainty in $\xi_{\psi}(\rho)$ to $O(1/\tau)$, which is clear from  Eq.~\eqref{eqn:wittness1}. 

The maximization over $\delta\rho$ becomes complicated due to the positivity constraint. Fortunately, the certification fails once we find some $\delta \rho$ such that the R.H.S. of Eq.~\eqref{eqn:main2} is exponential in $N$. A concrete example for this is provided by the GHZ state~\cite{greenberger2007}, defined as $\ket{\text{GHZ}}=\frac{1}{\sqrt{2}}\ket{00\ldots0}+\ket{11\ldots 1}$. 
  where we just choose $\delta \rho \propto |00\ldots0\rangle \langle 11\ldots1|+ |11\ldots1\rangle \langle 00\ldots0|=\otimes_i\sigma^+_i+\otimes_i\sigma^-_i$, which has an operator size of $N$. Since the lab state is fixed to be $P_{\psi}$ in Eq.~\eqref{eqn:main2}, we work in the large-$\alpha$ limit to obtain $\tau\left\rVert \tilde{\xi}_\psi-\xi_\psi\right\rVert_{\alpha}\geq (\sqrt{3/2})^N$. As a result, it's impossible to certify the GHZ state efficiently with local-Haar measurements.
  More generally, the denominator of Eq.~\eqref{eqn:main2} is dominated by contributions from operators with small size. Therefore, meaningful bounds that prevent efficient certification of quantum states can be obtained if one can find a $\delta \rho$ that contains only operators with a size close to $N$. In the supplementary material~\cite{SM}, we discussed a series of GHZ-like states, which exhibits global entanglement and can not be efficiently certified by arbitrary protocol based on local Haar measurements.

  \emph{ \color{blue}Discussions.--} In this work, we derive bounds for sample fluctuations in the task of quantum state certification using local Haar measurements. We first fix the (state-dependent) measurement observable, $\hat{\xi}_\psi$, and relate the sample fluctuations to a generating function for its size distribution, given by Eq.~\eqref{eqn:res_gen}. This provides insight into how efficiency varies across different protocols. As an example, we demonstrate that the certification of typical Haar random states is efficient for shadow overlap but not for direct fidelity estimation. We further generalize our analysis to arbitrary observables that can bound the fidelity, given by Eq.~\eqref{eqn:main2}. The results show that a class of GHZ-like states cannot be efficiently certified using local Haar measurements, regardless of the specific protocol. This reveals a general obstacle to pure state certification in many-body systems.

  Although we focus on setups with single-site random measurements, it is straightforward to generalize our analysis to scenarios where deterministic finite-depth quantum circuits, $\hat{U}_c$, are applied before random measurements. This can be mapped to the local certification problem with the lab state $\hat{U}_c\rho\hspace{.05em}\hat{U}_c^\dagger$ and the target state $\hat{U}_c|\psi\rangle$, allowing our bounds to be applied with minor adjustments. For random quantum circuits, the results contains additional ensemble average, which requires additional efforts. Another interesting question is to gain a deeper understanding of quantum states where the R.H.S. of Eq.~\eqref{eqn:main2} becomes exponential in $N$, possibly in relation to quantum phases and quantum error correction codes. We leave a detailed analysis of these questions for future study.

  \textit{Acknowledgments.}
  We thank Xiaodi Li and Liang Mao for helpful discussions. This project is supported by the NSFC under grant numbers 12374477.

\bibliography{Fidelity_Estimation.bbl}

\ifarXiv


\section*{Supplementary material}

# Supplementary Material: Bounding the Sample Fluctuation for Pure States Certification with Local Random Measurement

Langxuan Chen[1] and Pengfei Zhang[2, 3, 4, 5, *]

[1]*School of Physics, Xi'an Jiaotong University, Xi'an 710049, China*
[2]*Department of Physics, Fudan University, Shanghai, 200438, China*
[3]*State Key Laboratory of Surface Physics, Fudan University, Shanghai, 200438, China*
[4]*Shanghai Qi Zhi Institute, AI Tower, Xuhui District, Shanghai 200232, China*
[5]*Hefei National Laboratory, Hefei 230088, China*
(Dated: October 22, 2024)

### Equivalence of our protocol to measurements based on local Haar unitaries and the irrelevance of $\alpha_i$

In the main text, we based our analysis on projective measurements of each qubit alone a random direction $\boldsymbol{v}_i \in S^2$, where $i \in \{1, \ldots, N\}$ labels different qubits. On each qubit $i$, the observable to be measured is $(\boldsymbol{v}_i \cdot \sigma)_i$, where the subscript $i$ specifies the qubit it acts on. On the other hand, in classical shadow tomography [1–21], the lab state $\rho$ is first transformed by a random unitary operation $\rho \mapsto U\rho U^\dagger$ generated from some ensemble and then measured in the $Z$-basis. Measuring a state under $Z$-basis means that each qubit is measured by the operator $(\sigma_z)_i$, where $i$ specifies the qubit on which $\sigma_z$ acts. The data of the selected random unitary operator $U$ and the measured eigenstates are then recorded by a classical computer.

When the ensemble of unitary operators $U$ are chosen to be local Haar unitaries ("local twirling"), we have $U = \otimes_{i=1}^{N} u_i$, with $u_i$ being single qubit Haar random unitaries. As a result, we have

$$U^\dagger (\sigma_z)_i U = (u_i^\dagger \sigma_z u_i)_i = (\boldsymbol{v}_i \cdot \sigma)_i. \tag{1}$$

Here $\boldsymbol{v}_i \in S^2$ are independent random variables subject to a uniform probability distribution on the sphere $S^2$. This justifies the equivalence between our measurement protocols and the randomized measurements based on local Haar random unitaries used in classical shadow tomography. However, here we focus on the generic properties of any estimator constructed based on the data. While in classical shadow tomography they focus on the explicit constructions and properties of certain estimators.

As explained in the main text, the outcomes of our measurements include the final state $|\boldsymbol{n}\rangle \langle \boldsymbol{n}|$ of the qubits and the eigenvalues $\alpha_i$ of the operator $(\boldsymbol{v}_i \cdot \sigma)$. We argued that $\alpha_i$ contain no information of the state $\rho$ and hence might be ignored. Here we give the reason. Recall that the directions of measurement $\boldsymbol{v}_i$ are uniformly distributed random variables on the sphere, we have the joint probability distribution of $\alpha_i$ and $\boldsymbol{v}$:

$$p_\rho(\alpha, \boldsymbol{v}) = \frac{\langle \boldsymbol{n}| \rho |\boldsymbol{n}\rangle}{2^N}, \quad \sum_{\alpha_i = \pm 1} \int \frac{d\boldsymbol{v}}{(2\pi)^N} p_\rho(\alpha, \boldsymbol{v}) = 1, \tag{2}$$

where $\boldsymbol{n}_i = \alpha_i \boldsymbol{v}_i$ and $\frac{1}{2^N}$ are introduced for normalization. Change the variable to $\boldsymbol{n}$ we obtain

$$p_\rho(\alpha, \boldsymbol{n}) = \frac{\langle \boldsymbol{n}| \rho |\boldsymbol{n}\rangle}{2^N}. \tag{3}$$

Therefore, when $\boldsymbol{n}$ is given, the $2^N$ values of $\alpha_i$ occurs with equal probability, and thus contain no information of the qubits state $\rho$. As a result, only the statistics of $\boldsymbol{n}$ are of our interest, which is given by $P(\rho, \boldsymbol{n}) = \langle \boldsymbol{n}| \rho |\boldsymbol{n}\rangle$.

### Properties of the function $K_{\alpha,N}(\rho)$

When we discuss the variance bound in the main text, we introduced the function

$$K_{\alpha,N}(\rho) := \left[ \int \frac{d\boldsymbol{n}}{(2\pi)^N} \frac{1}{P(\rho, \boldsymbol{n})^\Delta} \right]^{\frac{1}{\Delta+1}}, \tag{4}$$

where we have $\Delta = \frac{2}{\alpha - 2}$. Here, we should discuss some properties of the function, together with a conjectured upper bound. From the definition, it's clear that when $\Delta \geq 1$ (or equivalently, $\alpha \leq 4$), the integral might diverge for those density matrix $\rho$ with singular eigenvalue.

As proposed in the main text, we should first prove that $K_{\alpha,N}(\rho)$ is minimized by $\rho_0 = \mathbb{I}/2^N$. Let's focus on the integral

$$F_\Delta(\rho) := \int \frac{\mathrm{d}\boldsymbol{n}}{(2\pi)^N} \frac{1}{P(\rho,\boldsymbol{n})^\Delta}. \tag{5}$$

By symmetry, we know that $F_\Delta(\rho)$ is invariant under local unitary transformation $\rho \to U\rho U^\dagger$, where $U = \otimes_i u_i$ are products of single qubit gates. What's more, being the integral of a convex function, $F_\Delta(\rho)$ itself is a convex function of $\rho$. It means that for positive constants $\lambda_i$ with $\sum_i \lambda_i = 1$, we have the inequality

$$\sum_i \lambda_i F_\Delta(\rho_i) \geq F_\Delta\left(\sum_i \lambda_i \rho_i\right). \tag{6}$$

For any density matrix $\rho$, we have

$$\rho_0 := \mathbb{I}/2^N = \int_{\text{Local Haar}} \mathrm{d}U \, U\rho U^\dagger. \tag{7}$$

Combine Eq. (6) and Eq. (7), we have

$$F_\Delta(\rho_0) \leq \int_{\text{Local Haar}} \mathrm{d}U \, F_\Delta(U\rho U^\dagger) = F_\Delta(\rho). \tag{8}$$

Therefore $F_\Delta(\rho)$ is minimized by $\rho_0 = \mathbb{I}/2^N$, for which we have $F_\Delta(\rho_0) = 2^{N(1+\Delta)}$. As a result, for $\alpha \in [2, \infty]$ we have

$$\min_\rho K_{\alpha,N}(\rho) = K_{\alpha,N}\left(\mathbb{I}/2^N\right) = 2^N. \tag{9}$$

On the other hand, the maximum of $K_{\alpha,N}(\rho)$ is harder to deal with. Being a convex function of $\rho$, we know that $F_\Delta(\rho)$ (thus $K_{\alpha,N}(\rho)$) must be maximized by some pure state. Setting $\rho = |\psi\rangle\langle\psi|$, we may differentiate $F_\Delta$ with respect to $\psi$ to obtain

$$\delta F_\Delta(\psi) = -\Delta \int \frac{\mathrm{d}\boldsymbol{n}}{(2\pi)^N} \frac{\langle\delta\psi|\boldsymbol{n}\rangle\langle\boldsymbol{n}|\psi\rangle + \langle\psi|\boldsymbol{n}\rangle\langle\boldsymbol{n}|\delta\psi\rangle}{|\langle\boldsymbol{n}|\psi\rangle|^{2(\Delta+1)}} = 0. \tag{10}$$

It's not easy to maximize $F$ through determine all solutions of Eq. (10). However, we may offer a reasonable conjecture. Notice that the main contribution of integral (5) comes from places where $|\langle\psi|\boldsymbol{n}\rangle|^2 = 0$. Thus, to maximize $F_\Delta(|\psi\rangle\langle\psi|)$, it's natural to choose those $\psi$ that are "most singular". A good choice is simply the product state $|\psi\rangle = |0\ldots00\rangle$, for which $\langle\psi|\boldsymbol{n}\rangle$ is singular whenever $\boldsymbol{n}_i = \hat{z}$ for some $i$. It's not hard to check that the choice $|\psi\rangle = |0\ldots00\rangle$ satisfies Eq. (10) by simple symmetry arguments. Therefore, we *conjecture* that $K_{\alpha,N}(\rho)$ is maximized by product states. For product states $\psi$,

$$F_\Delta(\psi) = \left(\int_{-1}^1 \mathrm{d}\cos\theta \left(\cos\frac{\theta}{2}\right)^{-2\Delta}\right)^N = \left(\frac{2}{1-\Delta}\right)^N. \tag{11}$$

Therefore, if the conjecture were true, we would have

$$\max_\rho K_{\alpha,N}(\rho) = \left(\frac{2}{1-\Delta}\right)^{\frac{N}{1+\Delta}} = (k_\alpha)^N, \tag{12}$$

where we have defined the constant $k_\alpha = \left(\frac{2(\alpha-2)}{\alpha-4}\right)^{1-2/\alpha}$. Note that the constant $k_\alpha$ is independent of $N$, and we have $k_\alpha \approx 2$ when $\alpha$ is large.

### Scaling of the variance bound with respect to sample size $M$

In the main text we only discussed the construction of the estimator $\tilde{\xi}(\boldsymbol{n})$ of $\xi(\rho)$ based on a *single* measurement outcome $\boldsymbol{n}$. For generic sample size $M$, the estimator $\tilde{\xi}^{(M)}$ is then assumed to be the *sample average* of the estimator

$\tilde{\xi}(\boldsymbol{n})$. When $\alpha$ is an even integer, it is possible to expand the power explicitly. Assuming the random variable $\tilde{\xi}$ has vanishing odd central moments, we have the bound

$$\left\| \frac{1}{M} \sum_{i=1}^{M} \tilde{\xi}(\boldsymbol{n}^{(i)}) - \xi(\rho) \right\|_\alpha \geq \frac{\|\tilde{\xi}(\boldsymbol{n}) - \xi(\rho)\|_\alpha}{M^{1-1/\alpha}}, \tag{13}$$

where $n^{(i)}$ is the outcomes of the $i$-th measurement. This is the bound proposed in the main text. However, there are several reasons to study more general scenarios. First it restricts us to the case $\alpha = 2n$, and also requires odd central moments like $\mathbb{E}[(\tilde{\xi} - \xi)^3]$ to vanish, which contradicts our purpose of keeping the discussion as general as possible. What's more, there are examples in probability theory when the best estimator is not a sample average. A simple example is provided by considering a random variable $X$ that is uniformly distributed on the interval $[\theta - \frac{1}{2}, \theta + \frac{1}{2}]$. For $M$ independent samples of $X$, the function $\tilde{\theta}(X^{(i)}) := \frac{1}{2}(\max_i(X^{(i)}) - \min_i(X^{(i)}))$ is clearly an unbiased estimator of $\theta$. When $M$ is large, the error of $\tilde{\theta}$ scales as $1/M$, thus outperforms any estimator that is constructed through a sample average.

To address the above problems, we would take $\tilde{\xi}^{(M)}$ and $\alpha$ to be general. The only requirement for $\tilde{\xi}^{(M)}$ is the expectation value condition $\mathbb{E}[\tilde{\xi}^{(M)}] = \text{tr}[\hat{\xi}\rho]$. Taking variation with respect to $\delta\rho$ as before we obtain

$$\sum_{i=1}^{M} \mathbb{E}\left[ (\delta P_i/P_i)(\tilde{\xi}^{(M)} - \xi) \right] = \text{tr}[\hat{\xi}\delta\rho], \tag{14}$$

where we have used the fact that for independent samples $\boldsymbol{n}^{(i)}$, the probability distribution $P^{(M)}(\rho, \boldsymbol{n}^{(i)})$ satisfies $P^{(M)}(\rho, \boldsymbol{n}^{(i)}) = \prod_{i=1}^{M} P(\rho, \boldsymbol{n}^{(i)})$. Applying generalized Hölder's inequality for fixed sample label $i$ we have the inequality

$$\left| \mathbb{E}\left[ (\delta P_i/P_i)(\tilde{\xi}^{(M)} - \xi) \right] \right| \leq \left\| \tilde{\xi}^{(M)} - \xi \right\|_\alpha \sqrt{K_{\alpha,N}(\rho) \times \left[ \int \frac{d\boldsymbol{n}}{(2\pi)^N} P(\delta\rho, \boldsymbol{n})^2 \right]}. \tag{15}$$

Since Eq. (15) is independent of $i$, we have

$$\left\| \tilde{\xi}^{(M)} - \xi \right\|_\alpha \geq \max_{\text{tr}[O]=0} \frac{M^{-1}|\text{tr}[\hat{\xi}O]|}{\sqrt{K_{\alpha,N}(\rho) \times \left[ \int \frac{d\boldsymbol{n}}{(2\pi)^N} P(O, \boldsymbol{n})^2 \right]}}, \tag{16}$$

where $O$ could be any traceless Hermitian operator. The result is clear: For generic sample size $M$, our bound simply scales as $1/M$ times the result for single-outcome-based estimators. It's interesting that the scaling here coincides with the ones found in our example. It is somehow suggesting that $1/M$ is the best error scaling one could obtain through increasing the sample size $M$. Note that the result is independent of the choice of $\alpha$ and the specific construction of the multi-sample estimator $\tilde{\xi}^{(M)}(\boldsymbol{n}^{(i)})$.

**Application to measurements of generic observables**

In this section, we discuss the implications of our bound to the measurement of a generic observable $\hat{\xi}$. As we have mentioned in the main text, the estimator-dependent variance bound is actually valid for any observable $\hat{\xi}$. In the main text, we kept the bound dependent on $K_{\alpha,N}(\rho)$ (hence on the lab state $\rho$) since the construction of observable-independent variance bounds for certification requires the choice $\rho = |\psi\rangle\langle\psi|$, where $|\psi\rangle$ is the pure state to be certified. However, for the measurement of an generic observable $\hat{\xi}$, no such requirement arise. As a result, we might maximize the bound over all possible lab state $\rho$ to obtain

$$\left\| \tilde{\xi}(\boldsymbol{n}) - \xi \right\|_\alpha \geq \max_{\text{tr}[O]=0} \frac{|\text{tr}[\hat{\xi}O]|}{\sqrt{2^N \times \left[ \int \frac{d\boldsymbol{n}}{(2\pi)^N} P(O, \boldsymbol{n})^2 \right]}}, \tag{17}$$

where the maximization have set $K_{\alpha,N}(\rho) = 2^N$. Maximize again over traceless Hermitian operator $O$ we obtain

$$\left\| \tilde{\xi} - \xi \right\|_\alpha^2 \geq \langle \hat{\xi}^2 \rangle \sum_{s>0} 3^s P_\xi(s), \tag{18}$$

where

$$\langle \hat{\xi}^2 \rangle = \text{tr}[\xi^2]/2^N, \quad \hat{\xi} = \sqrt{\langle \hat{\xi}^2 \rangle} \sum_P c_\xi(P) P. \tag{19}$$

In the main text we obtained a similar inequality through the substitution $\sum_{s>0} 3^s P_\xi(s) \geq \overline{3^s} - 1$, which is a minor modification when the size of $\xi$ is large. Note that since we have taken the maximization over $\rho$, It's suffix to have $\alpha \geq 2$. The above bound is valid for the estimation of any observable through local-Haar measurements. As an application, we may check how bound here compares with the result in classical shadow tomography. For simplicity, we shall analyze the estimation of Pauli string operators $\xi = P$. For a Pauli String operator of length $1 \leq k \leq N$, we have $\langle \hat{\xi}^2 \rangle = 1$ and $\sum_{s>0} 3^s P_\xi(s) = 3^k$. For simplicity we take $\alpha = 2$, which gives the variance bound:

$$\text{Var}(\tilde{P}(\boldsymbol{n})) \geq 3^k. \tag{20}$$

On the other-hand, we know that for classical shadow tomography, the estimators are explicitly constructed, with the variances bounded from above by the shadow norm $\|\hat{\xi}\|_{\text{sh}}^2$ [2]. For random local Haar measurements, the shadow norm of a Pauli string operator of length $k$ is simply given by $\|P\|_{\text{sh}}^2 = 3^k$, which *equals* to the lower bound obtained here. As a result, we conclude that when the randomized measurement is based on single qubit Haar unitaries, the estimators for Pauli string operators constructed in classical shadow tomography are in fact *optimal*.

### A review and comparison to shadow overlap

In this part we shall give a brief review about pure states certification through shadow overlap [22]. Since the original protocol was not based on local Haar measurements, we would propose a generalization of this formalism for local Haar measurements, so that we could compare our results.

Let's first describe the original construction of shadow overlap. Assuming the lab state $\rho$ is an $N$-qubit state, and $|\psi\rangle$ is a pure target state. In each measurement, one first choose $k \in \{1, \ldots, N\}$ and $O_k \in \{\sigma_x, \sigma_y, \sigma_z\}$ randomly. All qubits except the $k$-th are then measured in the Pauli $Z$-basis, and the $k$-th qubit is measured by $O_k$.

The measurement outcome is described by a string $\{z_i\} \in \{0,1\}^{N-1}$ (measurement outcomes for qubits with $i \neq k$) and $|s\rangle\langle s|$ (final state of the $k$-th qubit). Based on the outcomes, an estimator function $\omega_\psi$ is constructed as

$$\omega_\psi = \langle \psi_{k,z} | (3 |s\rangle\langle s| - 1) | \psi_{k,z} \rangle, \tag{21}$$

where

$$|\psi_{k,z}\rangle := \frac{\langle\{z_i\}|\psi\rangle}{\|\langle\{z_i\}|\psi\rangle\|}. \tag{22}$$

Note that since $|\{z_i\}\rangle$ is the state of all qubit except $k$, $|\psi_{k,z}\rangle$ is a normalized state in the Hilbert space of the $k$-th qubit. By construction, we always have $-1 \leq \omega_\psi \leq 2$ for any measurement outcomes. The central quantity of interest is the average $\mathbb{E}[\omega_\psi] = \text{tr}[\hat{L}_\psi \rho]$ of the estimator $\omega$. Since for random Pauli measurement on the $k$-th qubit we have $\mathbb{E}_{\rho_k}[3|s\rangle\langle s| - 1] = \rho_k$, the expectation therefore given by

$$\mathbb{E}[\omega_\psi] = \frac{1}{N} \sum_{k,\{z_i\}} \frac{\langle \psi | P_{k,\{z_i\}} \rho P_{k,\{z_i\}} | \psi \rangle}{\langle \psi | P_{k,\{z_i\}} | \psi \rangle}, \tag{23}$$

where we have defined the orthonormal projector $P_{k,\{z_i\}} = \otimes_{i \neq k} P_{z_i}$, with $P_{z_i}$ being the $Z$-basis single qubit projector acting on the $i$-th qubit. As a result, we have

$$\hat{L}_\psi = \frac{1}{N} \sum_{k,\{z_i\}} \frac{P_{k,\{z_i\}} |\psi\rangle \langle \psi| P_{k,\{z_i\}}}{\langle \psi | P_{k,\{z_i\}} | \psi \rangle}. \tag{24}$$

The matrix element in $Z$-basis is given by

$$\langle x | \hat{L}_\psi | y \rangle = \begin{cases} \frac{1}{N} \frac{\psi(x)\psi(y)^*}{|\psi(x)|^2 + |\psi(y)|^2} & x \sim y, \\ \frac{1}{N} \sum_{x' \sim x} \frac{|\psi(x)|^2}{|\psi(x)|^2 + |\psi(x')|^2} & x = y, \\ 0 & \text{otherwise,} \end{cases} \tag{25}$$

where $x, y \in \{0, 1\}^N$ labels the $Z$-basis vectors and $x \sim y$ when they differ exactly in one bit. $\psi(x) = \langle x|\psi\rangle$ are the wave functions of the pure state in $Z$-basis. It's not hard to see that $\hat{L}|\psi\rangle = |\psi\rangle$. Moreover, in [22] it was observed that $\hat{L}_\psi$ is related to the Markov chain with transition probability

$$P(x, y) = \begin{cases} \frac{1}{N} \frac{\pi(y)}{\pi(x)+\pi(y)} & x \sim y, \\ \frac{1}{N} \sum_{x' \sim x} \frac{\pi(x)}{\pi(x)+\pi(y)} & x = y, \\ 0 & \text{otherwise,} \end{cases} \qquad (26)$$

through an invertible linear transformation. Here $\pi(x) = |\psi(x)|^2$ are the probability distribution. As a result, $\hat{L}_\psi$ and $P(x, y)$ share the same spectrum. Since transitions are only allowed for $x \sim y$ in $P(x, y)$, it defines a random walk on the Hypercube $\{0, 1\}^N$.

Since $\hat{L}_\psi$ share the same spectrum with $P(x, y)$ and $\hat{L}_\psi$ is positively definite, its eigenvalues $\lambda_i$ satisfies $1 = \lambda_1 \geq \lambda_2 \geq \ldots \lambda_{2^N} \geq 0$. The *spectral gap* of $\hat{L}_\psi$ is then defined as $\Delta(\hat{L}_\psi) = 1 - \lambda_2$. On the other-hand, for the Markov chain $P$, the *relaxation time* $\tau$ is defined so that $\lambda_2 = 1 - \frac{1}{\tau}$. Therefore, we have $\Delta(\hat{L}_\psi) = \frac{1}{\tau}$.

The spectral gap of $\hat{L}_\psi$ plays a central role in the certification procedure based on shadow overlap. Let $|\psi_i\rangle$ be the eigenstates of $\hat{L}_\psi$ with eigenvalues $\lambda_i$, then

$$\langle\psi|\rho|\psi\rangle \leq \text{tr}[\hat{L}_\psi \rho] \leq 1 - \frac{1 - \langle\psi|\rho|\psi\rangle}{\tau}. \qquad (27)$$

The second inequality is true since

$$\text{tr}[\hat{L}_\psi \rho] = \langle\psi|\rho|\psi\rangle + \sum_{i \geq 2} \lambda_i \langle\psi_i|\rho|\psi_i\rangle \leq 1 - \frac{1}{\tau} \sum_{i \geq 2} \langle\psi_i|\rho|\psi_i\rangle. \qquad (28)$$

It should be clear that $\tau$ is the same proportionality constant we introduced in the observable independent bound discussed in the main text.

Since the estimator $\omega_\psi$ is bounded, the protocol would be able of certifying the state $|\psi\rangle$ efficiently if the relaxation time of the Markov chain satisfies $\tau \sim \text{poly}(N)$. In [22], it was proved based on the mathematical theory of Markov chain that as $N \to \infty$, $\tau \sim O(N^2)$ for most pure states drawn from a Haar ensemble, with the probability of exceptions being exponentially small for large $N$. However, it's worth mentioning that if the wave functions $\psi(x)$ vanishes for some values of $x$, the relaxation time $\tau$ could be infinite, resulting in a non-certifiable state. In addition, when $\pi(x), \pi(y)$ are both zero for some $x \sim y$, the denominator in Eq. (26) is zero, and a regularization is required.

Since the shadow overlap here is formulated by random local Pauli measurements in $X$, $Y$ and $Z$ basis, we should first generalize it to local Haar measurements before making comparison with our approach. The generalization is easy: Before each measurement, one may simply perform a unitary transformation $\rho \to U\rho U^\dagger$ on the lab state $\rho$, and then performing the same measurement protocol but evaluate the estimator $\omega_{U\psi}$ instead of $\omega_\psi$. The procedure here is intuitively clear since when $\rho$ and $|\psi\rangle\langle\psi|$ are close, $U\rho U^\dagger$ and $U|\psi\rangle\langle\psi|U^\dagger$ should also be close.

The resulting estimator based on the outcomes $|\boldsymbol{n}\rangle\langle\boldsymbol{n}| = \otimes_{i=1}^N |\boldsymbol{n}_i\rangle\langle\boldsymbol{n}_i|$ of local Haar measurements is defined as

$$\tilde{\Omega}_\psi(\boldsymbol{n}) = \frac{1}{N} \sum_{k=1}^N \frac{3\langle\boldsymbol{n}|\psi\rangle\langle\psi|\boldsymbol{n}\rangle}{\langle\psi|(\otimes_{i \neq k} |\boldsymbol{n}_i\rangle\langle\boldsymbol{n}_i|)|\psi\rangle} - 1, \qquad (29)$$

where we have replaced the random choice of $k$ by an average for a single sample $\boldsymbol{n}$, which do not change the expectation. Since the vector $\boldsymbol{n}$ is obtained in the measurement, the summation over $k$ might be performed on a classical computer. Like $\omega$, the estimator here also satisfies $-1 \leq \tilde{\Omega}_\psi(\boldsymbol{n}) \leq 2$. For generic lab state $\rho$, the expectation value of $\tilde{\Omega}_\psi$ is given by

$$\mathbb{E}[\tilde{\Omega}_\psi] = \int_{\text{Local Haar}} dU \, \text{tr}[U^\dagger \hat{L}_{U\psi} U \rho] = \text{tr}[\hat{\Omega}_\psi \rho], \qquad (30)$$

where $\hat{\Omega}_\psi = \int_{\text{Local Haar}} dU \, (U^\dagger \hat{L}_{U\psi} U)$ is the ensemble average of the original operator $\hat{L}_\psi$. Since we still have $\hat{\Omega}_\psi |\psi\rangle = |\psi\rangle$, the spectral gap of $\hat{\Omega}_\psi$ satisfies the inequality

$$\Delta(\hat{\Omega}_\psi) \geq \int_{\text{Local Haar}} dU \, \Delta(\hat{L}_{U\psi}). \qquad (31)$$

Note that set of unitaries $U$ for which the wave functions $\langle x| U |\psi\rangle$ vanishes for some $x \in \{0,1\}^N$ has zero measure in the local Haar ensemble. The fact should be clear from physicists' intuition, and could also be proved mathematically through a toy version of Sard's theorem (see e.g. [23]). As a result, no regularization is required for defining $\hat{\Omega}_\psi$ and all states $\psi$ are certifiable since the spectral gap $\Delta(\hat{\Omega}_\psi)$ is always strictly positive. In addition, since the global Haar ensemble is invariant under the action of local Haar unitaries $U$, the operator $\hat{\Omega}_\psi$ inherits the property from $\hat{L}_\psi$ that for most Haar states $\psi$ its inverse spectral gap scales as $\frac{1}{\Delta(\hat{\Omega}_\psi)} := \tau_\psi^\Omega \sim O(N^2)$, with probability of exceptions being exponentially small for large $N$. As a certification protocol based on random local Haar measurements, it subjects to the observable-independent bound we obtained in the main text. In the present case, we have $\|\tilde{\Omega}_\psi - \Omega_\psi\|_\alpha \leq 2$ for all $\alpha$ since $-1 \leq \tilde{\Omega}_\psi(\boldsymbol{n}) \leq 2$, which translates our bound to the inequality

$$2\tau_\psi^\Omega \geq \max_{P_\psi + \delta\rho \geq 0} \frac{|\mathrm{tr}[P_\psi \delta\rho]|}{\sqrt{2^N \int \frac{d\boldsymbol{n}}{(2\pi)^N} \langle\boldsymbol{n}|\delta\rho|\boldsymbol{n}\rangle^2}}, \tag{32}$$

where we have taken $\alpha \to \infty$. As an example, in the GHZ case the bound scales as $\tau_{\mathrm{GHZ}}^\Omega \gtrsim \left(\frac{3}{2}\right)^{N/2}$, which makes the efficient certification impossible.

### A discussion of generic states that are not efficiently certifiable

In the main text we proved that the GHZ states are not efficiently certifiable through local Haar measurements. Here we wish to discuss the generic features of states that are not efficiently certifiable. In the main text, we obtained the efficiency bound

$$\tau \left\|\tilde{\xi}_\psi - \xi_\psi\right\|_\alpha \geq \max_{P_\psi + \delta\rho \geq 0} \frac{|\mathrm{tr}[P_\psi \delta\rho]|}{\sqrt{2^N \int \frac{d\boldsymbol{n}}{(2\pi)^N} \langle\boldsymbol{n}|\delta\rho|\boldsymbol{n}\rangle^2}}, \tag{33}$$

for the certification procedure, where we assumed that $\alpha$ is large. Now, the squared denominator (denoted by $R(\delta\rho)$) might be expressed as

$$R(\delta\rho) = 2^N \int \frac{d\boldsymbol{n}}{(2\pi)^N} \langle\boldsymbol{n}|\delta\rho|\boldsymbol{n}\rangle^2 = \left(\frac{2}{3}\right)^N \sum_{S \subseteq \{1,\ldots,N\}} \mathrm{tr}\left[\delta\rho_S^2\right], \tag{34}$$

where $S = \{i_1, i_2, \ldots, i_k\}$ for $i_1 < i_2 \ldots < i_k \in \{1, \ldots, N\}$ represents a subcollection of the qubits, and $\delta\rho_S = \mathrm{tr}_{i \notin S}[\delta\rho]$ is the variation of the density matrix $\rho$ as observed in the Hilbert space of qubits within the subcollection $S$. Other qubits with $i \notin S$ are traced out. The length of $S$ is denoted by $|S| = k$.

For simplicity, we focus on the variation generated by differences between pure states $|\psi'\rangle$ and $|\psi\rangle$, that is, $\delta\rho = \rho' - \rho$, where $\rho = |\psi\rangle\langle\psi|$ and $\rho' = |\psi'\rangle\langle\psi'|$ are the pure state density matrix. This choice naturally preserves positivity. For generic $|\psi'\rangle$ that is not too close to $|\psi\rangle$, the numerator is of $O(1)$, leaving most parameter dependence in the denominator $R(\delta\rho)$, which we shall now investigate. For each choice of $S$, the contribution

$$\mathrm{tr}\left[\delta\rho_S^2\right] = \mathrm{tr}\left[(\rho_S - \rho_S')^2\right] \tag{35}$$

is the difference of the state $\psi$ and $\psi'$ measured by qubits in $S$. In particular, if $\psi$ and $\psi'$ looks the same for qubits in in $S$, we have $\rho_S = \rho_S'$, and $\mathrm{tr}\left[(\rho_S - \rho_S')^2\right] = 0$. Note that when this happens for some qubits collection $S$, it's also true for any subset (i.e. subsystem) of $S$.

The above discussion offers a natural way of finding non-efficient certifiable states $\psi$: one simply search for pairs of pure states (i.e. $\psi$ and $\psi'$) that looks the same once we trace out a few qubits in the system. Formally speaking, we define two states $\rho$ and $\rho'$ to be "equivalent up to $k$ qubits", if $\rho_S = \rho_S'$ for *all* qubit subcollections $S$ with $|S| \leq N-k$. In this case, the qubit state $\rho$ and $\rho'$ become equivalent once $k$ qubits are traced out. The relation is denoted $\rho \sim_k \rho'$. Note that $\rho \sim_0 \rho'$ means $\rho = \rho'$.

Giving a choice of $S$, we have $\mathrm{tr}[(\rho_S - \rho_S')^2] = \|\rho_S - \rho_S'\|_2^2 \leq 4$. As a result, for $\rho \sim_k \rho'$ and $k \leq N/2$, we have

$$R(\rho - \rho') = (2/3)^N \sum_{|S| > N-k} \mathrm{tr}[(\rho_S - \rho_S')^2] \leq 4k \times (2/3)^N \times \binom{N}{k}. \tag{36}$$

Taking the logarithm of $R$ we see that for large $N$, $1/R$ is exponential in $N$ (i.e. the state $\psi$ is not efficiently certifiable) if $k$ satisfies

$$\log 3/2 + \frac{k}{N} \log\left(\frac{k}{N}\right) + \left(1 - \frac{k}{N}\right) \log\left(1 - \frac{k}{N}\right) \geq 0, \tag{37}$$

which leads to the requirement $k/N \lesssim 0.1402$. We shall therefore try to find pairs of $\psi$, $\psi'$ so that $\rho_\psi \sim_k \rho_{\psi'}$ for relatively small values of $k$.

The physical reason behind this is intuitively clear. For states $\psi$ and $\psi'$, the relation $|\psi\rangle \sim_k |\psi'\rangle$ (for small values of $k$) implies that $|\psi\rangle$ and $|\psi'\rangle$ only differs in their *global* entanglement properties. To certify the state $|\psi\rangle$, we have to be able of distinguishing $|\psi'\rangle$ from $|\psi\rangle$. However, since $|\psi\rangle$ and $|\psi'\rangle$ only differs in global entanglement properties, they behave almost the same under local measurements, resulting in the failure of an efficient certification.

Since the case $k=0$ is trivial, we start by discussing inequivalent states $|\psi\rangle$ and $|\psi'\rangle$ with $\rho_\psi \sim_1 \rho_{\psi'}$. We will show that the requirement is very constraining and the only possible states are essentially the GHZ state up to some variations of coefficients and local unitary transformations.

Assuming the equivalence $\rho_\psi \sim_1 \rho_{\psi'}$ now, and consider the $i$-th qubit. We know that the pure states might be decomposed as

$$\begin{aligned} |\psi\rangle &= \sum_\alpha c_\alpha |\eta_\alpha\rangle \otimes |\psi_\alpha\rangle, \\ |\psi'\rangle &= \sum_\alpha c'_\alpha |\eta'_\alpha\rangle \otimes |\psi'_\alpha\rangle, \end{aligned} \tag{38}$$

where $\alpha = 0, 1$ and $|\eta_\alpha\rangle$, $|\eta'_\alpha\rangle$ are orthonormal basis in $\mathcal{H}_i$, the Hilbert space of the $i$-th qubit. Here $|\psi_\alpha\rangle$ and $|\psi'_\alpha\rangle$ are also orthonormal vectors in the Hilbert space of the remaining qubits. Decomposition (38) is essentially the *singular value decomposition* of the states. Normalization condition requires $\sum_\alpha |c_\alpha|^2 = \sum_\alpha |c'_\alpha|^2 = 1$. The equivalence of reduced density matrix after traced out $\mathcal{H}_i$ leads to

$$\sum_\alpha |c_\alpha|^2 |\psi_\alpha\rangle \langle \psi_\alpha| = \sum_\alpha |c'_\alpha|^2 |\psi'_\alpha\rangle \langle \psi'_\alpha|. \tag{39}$$

Therefore, we may redefine $c'_\alpha$ and $|\psi'_\alpha\rangle$ in the decomposition (38), so that $|\psi_\alpha\rangle = |\psi'_\alpha\rangle$ and $|c_\alpha|^2 = |c'_\alpha|^2$. The difference between $|\psi\rangle$ and $|\psi'\rangle$ then results from the difference of the orthonormal vectors $|\eta_\alpha\rangle$ and $|\eta'_\alpha\rangle$. We may then interpret the difference as a change of basis in $\mathcal{H}_i$, implemented by an unitary operator $U_i$ with

$$U_i = \sum_\alpha |\eta'_\alpha\rangle \langle \eta_\alpha|. \tag{40}$$

The operator $U_i$ is a local operator that acts on $\mathcal{H}_i$ alone. In addition, we have

$$U_i |\psi\rangle = \sum_\alpha c_\alpha U_i |\eta_\alpha\rangle \otimes |\psi_\alpha\rangle = \sum_\alpha c_\alpha |\eta'_\alpha\rangle \otimes |\psi_\alpha\rangle = |\psi'\rangle, \tag{41}$$

that is, $U_i$ takes the state $|\psi\rangle$ to $|\psi'\rangle$. In general, if we have $N$ Hilbert spaces $\mathcal{H}_i$ (not necessarily Hilbert space of a single qubit) and pairs of pure states $|\psi\rangle, |\psi'\rangle \in \mathcal{H} = \otimes_{i=1}^N \mathcal{H}_i$ such that for all $i$,

$$\text{tr}_i[\rho_\psi] = \text{tr}_i[\rho_{\psi'}]. \tag{42}$$

Then, for each choice of $i$, we may always find unitary operators $U_i$ acts on $\mathcal{H}_i$ with $U_i |\psi\rangle = |\psi'\rangle$. Decomposing $\mathcal{H}_i$ into the direct sums of eigenspaces of $U_i$, we get $\mathcal{H}_i = \oplus_\lambda V_\lambda^i$ for $V_\lambda^i$ being the eigenspace with eigenvalues $\lambda$. A generic pure state $|\psi\rangle \in \mathcal{H}$ might be decomposed as

$$|\psi\rangle = \sum_\lambda c_{\lambda_1,\cdots,\lambda_N}^{\alpha_1,\ldots,\alpha_N} \left|v_{\lambda_1}^{1,\alpha_1}\right\rangle \otimes \ldots \otimes \left|v_{\lambda_N}^{N,\alpha_N}\right\rangle, \tag{43}$$

where $\left|v_{\lambda_i}^{i,\alpha_i}\right\rangle$ are orthonormal basis vectors of the eigenspace $V_{\lambda_i}^i$, labeled by $\alpha_i$. Apply $U_i$ on $|\psi\rangle$, we obtain

$$U_i |\psi\rangle = \sum_\lambda \lambda_i c_{\lambda_1,\ldots,\lambda_N}^{\alpha_1,\ldots,\alpha_N} \left|v_{\lambda_1}^{1,\alpha_1}\right\rangle \otimes \ldots \otimes \left|v_{\lambda_N}^{N,\alpha_N}\right\rangle. \tag{44}$$

Since for any choice of $i, j$ we always have $U_i |\psi\rangle = U_j |\psi\rangle = |\psi'\rangle$, the coefficient function $c^{\alpha_1,\ldots,\alpha_N}_{\lambda_1,\ldots,\lambda_N}$ must satisfies the condition

$$\lambda_i c^{\alpha_1,\ldots,\alpha_N}_{\lambda_1,\ldots,\lambda_N} = \lambda_j c^{\alpha_1,\ldots,\alpha_N}_{\lambda_1,\ldots,\lambda_N}. \tag{45}$$

We thus conclude that $c^{\alpha_1,\ldots,\alpha_N}_{\lambda_1,\ldots,\lambda_N} = 0$ unless $\lambda_1 = \lambda_2 = \cdots = \lambda_N$. The resulting expansion for $|\psi\rangle$ is then simplified to

$$|\psi\rangle = \sum_\lambda c^{\alpha_1,\ldots,\alpha_N}_\lambda \left|v^{1,\alpha_1}_\lambda\right\rangle \otimes \ldots \otimes \left|v^{N,\alpha_N}_\lambda\right\rangle \tag{46}$$

where $\left|v^i_\lambda\right\rangle \in V^i_\lambda$ as before. Note that now all $\left|v^i_\lambda\right\rangle$ that appear in a single product have the same eigenvalue $\lambda$. In mathematical language, the above decomposition shows that $|\psi\rangle \in \oplus_\lambda \left(\otimes^N_{i=1} V^i_\lambda\right)$. In our analysis of $\rho_\psi \sim_1 \rho_{\psi'}$, we have $\dim \mathcal{H}_i = 2$. For $U_i$ to be non-trivial, each eigenspace $V^i_\lambda$ must then be one-dimensional, which makes the summation over $\alpha_i$ trivial. Eq. (46) then gives

$$|\psi\rangle = c_+ \left|v^1_+\right\rangle \otimes \ldots \otimes \left|v^N_+\right\rangle + c_- \left|v^1_-\right\rangle \otimes \ldots \otimes \left|v^N_-\right\rangle, \tag{47}$$

where $\left|v^i_+\right\rangle$ and $\left|v^i_-\right\rangle$ are orthonormal in the Hilbert space of the $i$-th qubit. The coefficient function satisfies $|c_+|^2 + |c_-|^2 = 1$ and $c_+ c_- \neq 0$. Define the local unitary transformation $W = \otimes^N_{i=1} W_i$ for $\left|v^i_+\right\rangle = W_i |1\rangle$, $\left|v^i_-\right\rangle = W_i |0\rangle$. Transform state $|\psi\rangle$ by $W$ we get

$$W^\dagger |\psi\rangle = c_+ |1\ldots 1\rangle + c_- |0\ldots 0\rangle. \tag{48}$$

In another word, for pure states $|\psi\rangle$ and $|\psi'\rangle$ with $\rho_\psi \sim_1 \rho_{\psi'}$ but $\rho_\psi \neq \rho_{\psi'}$, $|\psi\rangle$ and $|\psi'\rangle$ must be equivalent to states $c_+ |1\ldots 1\rangle + c_- |0\ldots 0\rangle$ through a local unitary transformation $W = \otimes^N_{i=1} W_i$. We shall refer to these states as generalized GHZ states. For the generalized GHZ states $|\psi\rangle = c_+ |1\ldots 1\rangle + c_- |0\ldots 0\rangle$ with $c_+, c_- \neq 0$, the observable-independent variance bound also scales as $(\sqrt{3/2})^N$ for large $N$, just like that for the GHZ state. One may check this by choosing $|\psi'\rangle = c_+ |1\ldots 1\rangle - c_- |0\ldots 0\rangle$, and set

$$\delta\rho = \rho_{\psi'} - \rho_\psi = 2\left(c_+ c^*_- |1\ldots 1\rangle\langle 0\ldots 0| + c_- c^*_+ |0\ldots 0\rangle\langle 1\ldots 1|\right). \tag{49}$$

The discussion for $k = 1$ shows that the only pure states satisfies $|\psi\rangle \sim_1 |\psi'\rangle$ are pairs of (generalized) GHZ states. In some sense, it makes them the *worst* possible target states for a local certification procedure.

On the other-hand, for larger values of $k$, it becomes much harder to determine all possible pairs of $|\psi\rangle$ and $|\psi'\rangle$ with $\rho_\psi \sim_k \rho_{\psi'}$. Since $\rho_\psi$ and $\rho_{\psi'}$ are only equivalent when $k$ qubits are traced out, we're lead to consider Hilbert spaces $\mathcal{H}_S = \otimes_{i \in S} \mathcal{H}_i$, where $S = \{i_1, \ldots, i_m\}$ labels a subsystem of the qubits. Following previous procedure, it's not hard to show that the condition $\rho_\psi \sim_k \rho_{\psi'}$ is equivalent to the requirement that for all $|S| \geq k$, there exist an unitary operator $U_S$ acts on the Hilbert space $\mathcal{H}_S$, such that $|\psi'\rangle = U_S |\psi\rangle$. For each $S$ with $|S| \geq k$, we have the eigenspace decomposition $\mathcal{H}_S = \oplus_\lambda V^S_\lambda$, where $V^S_\lambda$ are the eigenspace of the unitary operator $U_S$ that relates $|\psi\rangle$ and $|\psi'\rangle$. The expansion in Eq. (46) is still largely correct. However, instead of decomposing $\mathcal{H} = \otimes^N_{i=1} \mathcal{H}_i$, we work with the decomposition $\mathcal{H} = \otimes^m_{i=1} \mathcal{H}_{S_i}$, where $\{S_i\}^m_{i=1}$ are disjoint subcollections of the qubits $\{1, \ldots, N\}$ with $|S_i| \geq k$ and $\bigcup^m_{i=1} S_i = \{1, \ldots, N\}$. Mathematically speaking, $\{S_i\}^m_{i=1}$ forms a *partition* of $\{1, \ldots, N\}$. Giving a partition, the correct expansion for generic $k$ is given by

$$|\psi\rangle = \sum_{\lambda,\alpha} c^{\alpha_1,\ldots,\alpha_m}_{\{S_i\},\lambda} \left|v^{1,\alpha_1}_\lambda\right\rangle \otimes \ldots |v^{m,\alpha_m}_\lambda\rangle, \tag{50}$$

where $\left|v^{i,\alpha}_\lambda\right\rangle$ are orthonormal basis of the space $V^{S_i}_\lambda$. However, unlike the $k = 1$ case, we have inequivalent ways of *partitioning* $\{1, \ldots, N\}$ into unions of $S_i$. Since each partition leads to a different expansion, one has to check that these expansions are all *consistent* with each other. This is the most complicated part of solving the relation, and we haven't found a general way of determine all possible solutions. However, we still have simple solutions formed by GHZ-like states. For example, the state

$$|\psi\rangle = c_+ |\alpha\rangle \otimes |1, \ldots, 1\rangle + c_- |\beta\rangle \otimes |0, \ldots, 0\rangle, \tag{51}$$

where $|\alpha\rangle, |\beta\rangle$ are arbitrary normalized states in the Hilbert space of the first $m$ qubits is consistent with Eq. (50) for $k \geq m + 1$. One may also check that the state

$$|\psi'\rangle = c_+ |\alpha\rangle \otimes |1, \ldots, 1\rangle - c_- |\beta\rangle \otimes |0, \ldots, 0\rangle \tag{52}$$

satisfies $|\psi'\rangle \sim_k |\psi\rangle$ for all $k \geq m+1$, which makes it impossible to be certified efficiently through local Haar measurements when $m$ is small.

---


\fi

\end{document}